\documentclass[twocolumn]{article}

\usepackage{amsmath, amssymb, amsfonts}
	\numberwithin{equation}{section}
\usepackage[left=2cm,right=2cm,top=2.5cm,bottom=2.5cm]{geometry}
\usepackage{cite}
\usepackage{graphicx}
\usepackage{xcolor}
\usepackage{textcomp}
\usepackage{booktabs}
\usepackage{gensymb}
\usepackage{makeidx}
\usepackage{hyperref}
\usepackage{multicol}
\usepackage{multirow}

\usepackage{tikz}
\usetikzlibrary{calc}

\usepackage{outlines}
\usepackage{caption}
\usepackage{subcaption}
\usepackage{titlesec}
\usepackage{balance}
\usepackage{abstract}
\usepackage{titling}
\usepackage{authblk}
\usepackage{natbib}
\usepackage{har2nat}

\hypersetup{pdfborder={0 0 0},
	pdfusetitle,
	pdfauthor={Alberto Garinei et al.}}

\pretitle{
	\begin{center}
		\Huge\scshape 
	}
\posttitle{%
	\end{center}%
	\noindent\vrule height 1 pt width \textwidth
	\vskip .75em plus .25em minus .25em
}

\preauthor{\begin{flushleft}}
\postauthor{\par\end{flushleft}}

\newsavebox\affbox

\titleformat{\section}
{\scshape\large\filcenter}{\thesection.}{.7em}{}

\titleformat{\subsection}
{\normalfont\normalsize\bfseries\itshape}{\thesubsection}{1em}{}

\titleformat{\subsubsection}
{\normalfont\normalsize\itshape}{\thesubsubsection}{1em}{}

\tikzstyle{mybox} = [draw=black, 
rectangle, rounded corners, inner sep=10pt, inner ysep=20pt]
\tikzstyle{fancytitle} =[draw=black, fill=white, 
text=black, rounded corners, inner xsep=10pt, inner ysep=4pt]

\providecommand{\keywords}[1]
{
	\vskip 1em
	\noindent\rule[.3em]{3.5cm}{.7pt}
	
 	\small	
  	\noindent\textbf{\textit{Keywords}:} #1
}

\title{R\&D evaluation methodology based on group-AHP with uncertainty}

\author[1,*]{Alberto Garinei}
\author[2]{Emanuele Piccioni}
\author[3]{Massimiliano Proietti}
\author[3]{Andrea Marini}
\author[3]{Stefano Speziali}
\author[1]{Marcello Marconi}
\author[4]{Raffaella Di Sante}
\author[5]{Sara Casaccia}
\author[5]{Paolo Castellini}
\author[5]{Milena Martarelli}
\author[5]{Nicola Paone}
\author[5]{Gian Marco Revel}
\author[5]{Lorenzo Scalise}
\author[6]{Marco Arnesano}
\author[7]{Paolo Chiariotti}
\author[8]{Roberto Montanini}
\author[8]{Antonino Quattrocchi}
\author[9]{Sergio Silvestri}
\author[10]{Giorgio Ficco}
\author[11]{Emanuele Rizzuto}
\author[12]{Andrea Scorza}
\author[13]{Matteo Lancini}
\author[2]{Gianluca Rossi}
\author[2]{Roberto Marsili}
\author[7]{Emanuele Zappa}
\author[12]{Salvatore Sciuto}
\author[14]{Gaetano Vacca}
\author[14]{Laura Fabbiano}

\affil[1]{Department of Engineering Sciences, Guglielmo Marconi University, Rome, Italy}
\affil[2]{Department of Engineering, University of Perugia, Perugia, Italy}
\affil[3]{Idea-re S.r.l., Perugia, Italy}
\affil[4]{Department of Industrial Engineering-DIN, University of Bologna, Forlì, Italy}
\affil[5]{Università Politecnica delle Marche, Dipartimento di Ingegneria Industriale e Scienze matematiche (DIISM), Ancona, Italy}
\affil[6]{Università Telematica eCampus, Novedrate (CO), Italy}
\affil[7]{Department of Mechanical Engineering, Politecnico di Milano, Milan, Italy}
\affil[8]{Department of Engineering University of Messina, Messina, Italy}
\affil[9]{Research Unit of Measurements and Biomedical Instrumentation, Campus Bio-Medico University of Rome,  Rome, Italy}
\affil[10]{Department of Civil and Mechanical Engineering (DICEM), University of Cassino and Lazio Meridionale, Cassino (FR), Italy}
\affil[11]{Department of Mechanical and Aerospace Engineering, Sapienza, University of Rome, Rome, Italy}
\affil[12]{Department of Engineering, University of Roma Tre, Rome, Italy}
\affil[13]{Department of Mechanical and Industrial Engineering, University of Brescia, Brescia, Italy}
\affil[14]{Department of Mechanics, Mathematics and Management, Polytechnic University of Bari, Italy}
\affil[*]{Corresponding author: a.garinei@unimarconi.it}

\date{}

\begin{document}
\bibliographystyle{agsm}

\newgeometry{margin=2.5cm}
\begin{titlingpage}
\maketitle

  \begin{abstract}
  \normalsize\noindent In this paper, we present an approach to evaluate Research \& Development (R\&D) performance based on the Analytic Hierarchy Process (AHP) method. Through a set of questionnaires submitted to a team of experts, we single out a set of indicators needed for R\&D performance evaluation. The indicators, together with the corresponding criteria, form the basic hierarchical structure of the AHP method.  The numerical values associated with all the indicators are then used to assign a score to a given R\&D project. In order to aggregate consistently the values taken on by the different indicators, we operate on them so that they are mapped to dimensionless quantities lying in a unit interval. This is achieved by employing the empirical Cumulative Density Function (CDF) for each of the indicators. We give a thorough discussion on how to assign a score to an R\&D project along with the corresponding uncertainty due to possible inconsistencies of the decision process. A particular example of R\&D performance is finally considered.
  
  %
  \end{abstract}

\keywords{AHP, Multi-Criteria Decision Making, R\&D performance, R\&D measures}
\end{titlingpage}


\restoregeometry

\section{Introduction}

The Analytic Hierarchy Process (AHP) is a Multi-Criteria Decision Making (MCDM) method developed by Saaty in the 1970's (\citeasnoun{saaty1977scaling}).  It provides a systematic approach to quantifying relative weights of decision criteria. 

Its strength relies on the fact that it allows to decompose a decision problem into a hierarchy of sub-problems, each of which can be analyzed independently in a similar manner.  It is used in a wide variety of decision situations, in fields like education, industry, healthcare and so on.

In this paper, we propose a method to evaluate Research and Development (R\&D) performance, based on group-AHP, through the introduction of a ``score'' assigned to each R\&D projects in a given set. 

R\&D represents the set of innovative activities undertaken by companies and/or governments to develop new and more efficient services or products as well as to improve the existing ones. It has become somewhat crucial to have a systematic method to evaluate the performance a given project or research activity (\citeasnoun{lazzarotti2011model}).  See, among others, also (\citeasnoun{kerssens1999r}),  (\citeasnoun{moncada2010does}), (\citeasnoun{tidd2000managing}),  (\citeasnoun{griffin1997pdma}), (\citeasnoun{bremser2004utilizing}), (\citeasnoun{jefferson2006r}), (\citeasnoun{kim2002economic}), (\citeasnoun{kaplan1996using}), (\citeasnoun{chiesa2009performance}) and references therein for the importance of R\&D performance assessment. Quantitative methods coupled with qualitative assessments are used in decision support systems, for example by project funding commissions. 

However, there are currently no standards for measuring the performance of an R\&D project.  The method developed in this paper stems from a critical approach to the measurement problem concerning complex systems (such as Research and Development).  With the help of group multi-criteria methodologies,  we tried to faithfully represent the evaluations of R\&D projects through the involvement of stakeholders.  As a matter of fact, the latter represent diverse interests, and belong to different domains of knowledge.

We used three questionnaires addressed to stakeholders at different stages of the process with the ideal goal of developing a shared decision support tool that is easy to use and whose operation can be directly explained.
In view of adopting the logic of the metrological method, we defined a model capturing the subtle features of R\&D performance evaluation and
keeping track of measurements uncertainties.

In order to introduce the standard AHP decision structure, we need to define precisely what our criteria and sub-criteria will be. Criteria (or perspective in our parlance) are selected following the existing literature and,  more in detail,  have been identified to be: Internal Business perspective,  Innovation and Learning perspective, Financial perspective, Network and Alliances perspective.

We then single out a set of sub-criteria (indicators) through a set of questionnaires submitted to a team of experts selected from academia or private research hubs in Italy. The indicators, along with the corresponding criteria, will form in our analysis the basic hierarchical structure of the AHP method.

In order to have a sensible way to aggregate the values the different indicators take on, we operate on them in such way they share the same scale, namely they are all dimensionless quantities varying over the same range, which for convenience we choose to be 0 to 1. This is attained by employing as transformation map for each the indicator the corresponding empirical Cumulative Density Function (CDF). In this way, all the resulting variables are approximately uniformly distributed over the unit interval.

It is well-known that decision processes in complex systems carry along judgmental inconsistencies.  Aware of the fact that some inconsistencies are difficult to get rid of, we propose a rigorous method to quantify the uncertainty affecting the ``score'' of a given R\&D project. In order to better show how our method works, we give an example of application in the last section of this paper.
The method has been employed to evaluation of R\&D projects whose data are stored in the DPR\&DI (Digital Platform for R\&D and Innovation Projects).

This paper is organized as follows.  In Section \ref{Basics of the AHP method} we discuss in detail the basics of the AHP method as developed originally. In Section \ref{methodology} we propose a method to choose the criteria and sub-criteria to evaluate R\&D performance through a set of questionnaires.  We then give a detailed and precise account on how to evaluate R\&D performance of a given project, and finally we discuss the consistency of the proposed method. We give an example of R\&D performance evaluation in Section \ref{case study} and our conclusions in Section \ref{conclusions}.

\section{Theoretical Background: the AHP method}\label{Basics of the AHP method}

In this section, we discuss the basics of the AHP (Analytic Hierarchy Problem) method as developed originally by Saaty in the 1970's.  More details can be found,  for example, in the book (\citeasnoun{saaty2010mathematical}) or in the review (\citeasnoun{ishizaka2011review}). 

\subsection{Decision problems}

We face many decision problems in our daily lives.  They can be as simple as deciding what jeans we want to buy or more involved, like what person to hire for a post-doc position.  Whatever decision problem we are facing,  a systematic way to deal with it can be useful, and this is where AHP comes to play a role.

In AHP, each decision problem can be broken down in three components, each with the same basic structure:
\begin{itemize}
\item The \textit{goal} of the problem, namely the objective that drives the decision problem. 

\item The \textit{alternatives}, namely the different options that are being considered in the decision problem.

\item The \textit{criteria}, namely the factors that are used to evaluate the alternatives with respect to the goal.
\end{itemize}

Moreover, if the problem requires it, we can associate sub-criteria to each criterion, adding extra layers of complexity.  We will see an example of this in Section \ref{case study}. 

The three levels (or more if we consider sub-criteria) define a \textit{hierarchy} for the problem, and each level can be dealt with in a similar fashion to the others. This is essentially the basic structure of the AHP method in decision problems.  The rest of this section is devoted to spelling out the details of how a decision is eventually made.

\subsection{Weighting the problem}

A crucial ingredient in any decision problem is the mapping of notions, rankings etc. to numerical values.  Basic examples of mappings are scales of measurements, like the Celsius-degree for the temperature or dollars for money. In these cases we have what are called \textit{standard scales}, where standard units are employed to determine the \textit{weight} of an object. 

However,  it often happens that the same number (say 100\textdegree) means different things to different people, according to the situation, or different numbers are as good (or as bad) for a given purpose (e.g. when trying to find the right temperature for a fridge 100\textdegree\ is as bad as $-100$\textdegree). Moreover, it might be the case that we need to analyze processes for which there is no standard scale.  Thus, we need to find a way to deal with these situations consistently.

It turns out that what really matters is \textit{pairwise} comparisons between different options.  In this way we can create a relative ratio scale and, in fact, here is the crux of the AHP method, as we will see in a moment.

In the case we are dealing with a standard scale, we can assign to $n$ objects $n$ weights $w_1$, $\dots$, $w_n$.  Then, we can create a matrix\footnote{In this paper we deal mainly with finite dimensional real vector spaces.  In particular, if $V$ and $W$ are vector spaces of dimensions $n$ and $m$ respectively,  a choice of bases $v = \{v_1, \dots, v_n\}$ and $w = \{w_1, \dots, w_m\}$ determines isomorphisms of $V$ and $W$ with $\mathbb{R}^n$ and $\mathbb{R}^m$, respectively. Any linear operator from $V$ to $W$ has a matrix presentation $A \in \mathbb{R}^{n \times m}$ with respect to the given bases. In this respect, the eigenvalue eqn. \eqref{eigenvalue equation example} is a linear transformation from a space to itself.} $A \in \mathbb{R}^{n \times n}$ of pairwise comparisons in the following way
\begin{equation} \label{reciprocal matrix}
	A=\begin{pmatrix}
	w_1/w_1 & w_1/w_2 & \cdots & w_1/w_n \\ 
	w_2/w_1 & w_2/w_2 & \cdots & w_2/w_n\\ 
	\vdots & \vdots & \ddots & \vdots  \\ 
	w_n/w_1 & w_n/w_2 & \cdots & w_n/w_n\\ 
	\end{pmatrix}\, .
\end{equation}
The matrix $A$ is an example of a \textit{reciprocal} matrix, i.e. a matrix where each entry satisfies $a_{ij} = 1/a_{ji}$. This is indeed what we would expect when there is an underlying standard scale.  For example, if we are are to determine which among two apples is the reddest and, according to a given scale, apple $a$ is twice as red as apple $b$, it necessarily follows that apple $b$ is one-half as red as apple $a$.

Note the following interesting fact, that will be relevant for us later. If we define the vector $w = (w_1, \dots , w_n)^T$ it is easily seen that
\begin{equation}\label{eigenvalue equation example}
A \cdot w = n w \, ,
\end{equation}
where the dot-product is just matrix product, i.e. $w$ is an eigenvector of $A$ with eigenvalue $n$.  In fact, it is rather easy to convince ourselves that the matrix $A$ in eqn. \eqref{reciprocal matrix} has rank 1 and a theorem in linear algebra tells us that it must have only one non-zero eigenvalue.  On the other hand, the trace of a matrix gives the sum of it eigenvalues which in our case turns out to be $1 + \dots + 1 = n$.  It is therefore coherent to conclude that a consistent matrix like $A$ above has only one non-zero eigenvalue, $n$.  In this case $n$ is also called the \textit{principal eigenvalue}, i.e. the largest of the eigenvalues of a square matrix. 

As we said before, sometimes we have to deal with decision processes where a standard scale does not exist and thus we are not given a priori a weight vector $w$. What is really meaningful in this case is the matrix of pairwise comparisons between alternatives, similar to that in eqn. \eqref{reciprocal matrix}
\begin{equation} \label{reciprocal matrix no standard scale}
	A = (a_{ij})=\begin{pmatrix}
	1 & a_{12} & \cdots & a_{1n} \\ 
	a_{21} & 1 & \cdots & a_{2n}\\ 
	\vdots & \vdots & \ddots & \vdots  \\ 
	a_{n1} & a_{n2} & \cdots & 1\\ 
	\end{pmatrix}\, .
\end{equation}
Here $a_{ij}$ tells us how the $i$-th object compares to the $j$-th object according to a give criterion/goal.  Notice that also in this case we should impose $a_{ij} = 1/a_{ji}$, i.e. we should have a reciprocal matrix, but now each entry is not given by a ratio of two quantities. 

In order to make the pairwise-comparison coefficients $a_{ij}$ as explicit as possible, the Saaty's 1-9 scale is often used (see Figure \ref{Saaty scale}).
The scale should be read in the following way: If an object $i$ is as important as the object $j$, then we should set $a_{ij} = 1$. 
If, instead object $i$ is more important than the object $j$, then $a_{ij}$ should be set to $3$, $5$, $7$ or $9$, following the scheme in Figure \ref{Saaty scale}.
Also the intermediate even values (2, 4, 6, 8) can be used and allow for finer assessments.

\newcommand*{\TickSizeOdd}{3pt}%
\newcommand*{\TickSizeEven}{2pt}%
\begin{figure}[ht]
\begin{center}
		\begin{tikzpicture}[scale=0.5]
		\draw[->] (0,0) -- (0,10) node[above=5pt, midway, sloped] {Level of importance};
		\draw ($(0,1) + (-\TickSizeOdd,0)$) -- ($(0,1) + (\TickSizeOdd,0)$)
			node [right] {$1$ \, $\leftarrow$ \, Equal importance };
		\draw ($(0,3) + (-\TickSizeOdd,0)$) -- ($(0,3) + (\TickSizeOdd,0)$)
			node [right] {$3$ \, $\leftarrow$ \,   Moderate importance };
		\draw ($(0,5) + (-\TickSizeOdd,0)$) -- ($(0,5) + (\TickSizeOdd,0)$)
		node [right] {$5$ \, $\leftarrow$ \,  Essential or strong importance };
		\draw ($(0,7) + (-\TickSizeOdd,0)$) -- ($(0,7) + (\TickSizeOdd,0)$)
		node [right] {$7$ \, $\leftarrow$ \,  Very strong importance  };
		\draw ($(0,9) + (-\TickSizeOdd,0)$) -- ($(0,9) + (\TickSizeOdd,0)$)
		node [right] {$9$ \, $\leftarrow$ \,  Extreme importance };
		\foreach \y in {2,4,...,8} {%
			\draw[gray] ($(0,\y) + (-\TickSizeEven,0)$) -- ($(0,\y) + (\TickSizeEven,0)$)
			node [right, gray] {$\y$};
		}
	\end{tikzpicture}
\end{center}
\caption{Saaty's 1-9 scale.}
\label{Saaty scale}
\end{figure}
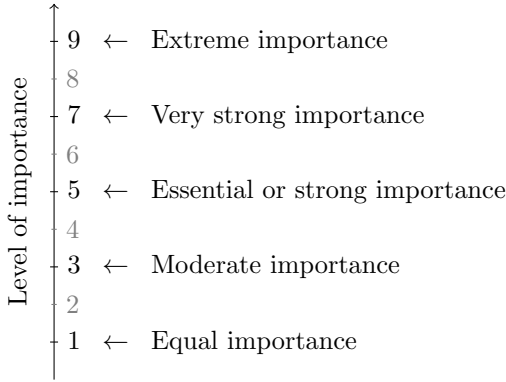


What if we considered an eigenvalue equation also for the matrix $A$ defined in \eqref{reciprocal matrix no standard scale}? And what would be the meaning of the weights (priorities) $w_i$ in this case? Let us begin by answering the first question first. 

The Perron-Frobenius theorem tells us that there exists one principal eigenvalue, $\lambda_{\text{max}}$, and that it is unique. We then find an equation of the form
\begin{equation}\label{eigenvalue equation}
\begin{pmatrix}
	1 & a_{12} & \cdots & a_{1n} \\ 
	a_{21} & 1 & \cdots & a_{2n}\\ 
	\vdots & \vdots & \ddots & \vdots  \\ 
	a_{n1} & a_{n2} & \cdots & 1\\ 
	\end{pmatrix}  \begin{pmatrix}
	w_1  \\ 
	w_2 \\ 
	\vdots \\ 
	w_n \\ 
	\end{pmatrix} = 
	\lambda_{\text{max}} \begin{pmatrix}
	w_1  \\ 
	w_2 \\ 
	\vdots \\ 
	w_n \\ 
	\end{pmatrix}
\end{equation}

It is a theorem (\citeasnoun{saaty1990make}) that for a reciprocal $n \times n$ matrix with all entries greater than zero, the principal eigenvalue $\lambda_{\text{max}}$ is always greater or equal to $n$,  $\lambda_{\text{max}} \ge n$.  In particular, $\lambda_{\text{max}} = n$ if and only if $A$ is a \textit{consistent matrix}. 

What is it meant by consistent matrix? If we reckon that alternative $i$ is $a_{ij}$ times better than alternative $j$, and the latter is $a_{jk}$ times better than alternative $k$, we should have, for consistency, $a_{ik} = a_{ij} a_{jk}$.  This is know as \textit{multiplicative consistency}. It is easily seen that multiplicative consistency implies reciprocity, but the converse is not true. 

It is often the case that multiplicative consistency is not respected, introducing some form of inconsistency in the evaluation process. One major drawback, for example, is that the fundamental scale ranges from $1/9$ to $9$ and a product of the form $a_{ij} a_{jk}$ might very well be outside the scale, making it impossible to respect multiplicative consistency.\footnote{There are different approaches to deal with the problem of the scale range. One approach could be to change the linear scale given before to a more convoluted one. For example in (\citeasnoun{donegan1992new}) an asymptotic scale is employed so that we never get out of a prefixed scale range.  However,  in the literature, the linear scale of Saaty seems to be the most widely used scale. } In the next subsection, we will see how to manage possible inconsistencies.

In order to have a (nearly) consistent matrix of pairwise comparisons $A$,  $\lambda_{\text{max}}$ should not differ much from the dimension of $A$, $n$.  In particular, finding the eigenvector $w = (w_1, \dots , w_n)^T$ amounts to finding the weights (or priorities) of the $n$ objects (alternatives), and we are assured that, if the matrix $A$ is sufficiently consistent, $a_{ij} \approx w_i/w_j$. Note that multiplying both sides of eqn.  \eqref{eigenvalue equation} by an arbitrary constant is harmless, and therefore the vector $w$ can be conveniently normalized as we please.  We will have to say a little more on this below. 

\subsection{How to compute weights}

We now find ourselves in the position where we should determine the priority vector $w$, eqn. \eqref{eigenvalue equation}, once a pairwise comparison matrix is given. The easiest way to do so is to solve eqn. \eqref{eigenvalue equation} using standard methods in linear algebra.  However, general procedures are not always exempt from inconsistencies (in AHP). For example, for inconsistent matrices with dimension greater than 3,  there is a right-left asymmetry, i.e a right-eigenvector is not a left-eigenvector. 

In order to avoid this issue, a common alternative to compute the priority vector $w$ makes use of the logarithmic least squares (LLS) method  
(\citeasnoun{de1984statistical}), (\citeasnoun{crawford1985note}). 
The relation between the matrix pairwise comparison $A$ and the relative priority vector $w$ can be expressed as
\begin{equation}\label{eq:perturbation_factor}
a_{ij} = \frac{w_i}{w_j} \, \varepsilon_{ij} \, , \qquad i,j = 1, \dots, n \, ,
\end{equation}
where $\varepsilon_{ij}$ are positive random perturbations.
It is commonly accepted that for nearly consistent matrices the $\varepsilon_{ij}$ factor is log-normal 
distributed\footnote{Indeed, the authors of (\citeasnoun{shrestha1991statistical}) found out that the error factors $\varepsilon_{ij}$ 
	best describe the inconsistency in the decision process when they are log-normal distributed
	\begin{equation}
	\log \varepsilon_{ij} \sim \mathcal{N}(0, \sigma_{ij}^2) \, ,
	\end{equation}
	where $\mathcal{N}(\mu, \sigma^2)$ is the normal Gaussian distribution function with mean $\mu$ and variance $\sigma^2$.
	In particular, note that the mean value of the error factor $\varepsilon_{ij}$ is 1 and its range can be varied by choosing $\sigma_{ij}^2$ accordingly with the degree of expertise.}. 
Thus, to determine the weights $w_i	$ one can take the logarithm of \eqref{eq:perturbation_factor} and then apply the least square principle, 
namely minimizing the sum of squares of $\log \varepsilon_{ij}$, 
\begin{equation}\label{log error}
	E(w) = \sum_{i,j = 1}^n \left( \log a_{ij} - \log(w_i) + \log(w_j) \right)^2 \, .
\end{equation}
An easy computation reveals that $E(w)$ is minimized when
\begin{equation}\label{geometric mean}
w_i = \left( \prod_{j = 1}^n a_{ij} \right)^{\frac{1}{n}}  \, , \qquad i = 1, \dots, n \, .
\end{equation}
This is also called the geometric mean, and from now on we will adopt this method to compute weights.  Note that for consistent matrices, $w_i$ as in \eqref{geometric mean} is an eigenvector with eigenvalue $n$. The weights $w_i$ in eqn. \eqref{geometric mean} are defined up to a multiplicative constant (see eqn. \eqref{log error}). We have normalized them so that $\prod_{j = 1}^n w_{j} = 1$. 

\subsection{Aggregation}

The final step is to aggregate local priorities across all criteria to in order to determine the global priority of each alternative. This step is necessary to determine which alternative will be the preferred one.

In the original formulation of AHP, this is done in the following way. If we denote $l_{ij}$ the local priority (weight) of the alternative $i$ with respect to the criterion $j$ and $w_j$ the weight of the criterion $j$, the global priority $p_i$ for the alternative $i$ is defined to be
\begin{equation}
p_i = \sum_j w_j l_{ij} \, .
\end{equation}

Criterion weights and local priorities can be normalized so that they sum up to 1. In this way, we find $\sum_i p_i = 1$. The alternative getting the highest priority (modulo inconsistencies to be discussed later) will be the favorite one in the decision process.

Let us now move on to discussing (some of the) possible inconsistencies of the AHP method.

\subsection{Consistency of the AHP method}

As we remarked before, the AHP method is based on the idea that there is always some underlying scale in a decision problem.  This is encoded in the fact that when we have calculated our weight matrix -- which by definition is a consistent ratio matrix built out of the weight ratios -- this one should not be too far off the original pairwise comparison matrix.

In order to determine how far off we are,  we need to find a way to determine the inconsistency of our decision matrices. To this purpose, it is useful to recall a couple of facts (\citeasnoun{saaty1990make}).  Saaty noticed that for a reciprocal $n \times n$ matrix $A$ with all entries bigger than zero, the principal eigenvalue is always equal or greater than $n$. This is easily proved with some simple linear algebra.

Moreover,  it turns out that $A$ is a fully consistent matrix if and only if the principal eigenvalue is strictly equal to $n$. 

Given these facts, it is possible to define a set of indices to measure the consistency of our decision matrices. In particular, we can define the \textit{Consistency Index} (CI) as
\begin{equation}
\text{CI} = \frac{\lambda_{\text{max}} - n}{n - 1} \, .
\end{equation}
Note that $\text{CI} \geq 0$, as a consequence of what we said above. Also, the more CI is different from zero the more inconsistent we have been in the decision process.

We can also define the \textit{Random Index} RI of size $n$ as the average CI calculated from a large number of randomly filled matrices. For a discussion on how these matrices are created see (\citeasnoun{alonso2006consistency}).  

Finally, we define the \textit{Consistency Ratio} CR as the ratio CI$(A)/$RI$(A)$ for a reciprocal $n \times n$ matrix, where RI$(A)$ is the random index for matrices of size $n$.

Usually, if the CR is less than 10$\%$ the matrix is considered to have an acceptable consistency.  Nonetheless, this consistency index is sometimes criticized as it allows contradictory judgments. See the review (\citeasnoun{ishizaka2011review}) for a discussion about this.

In the literature, several other methods to measure consistency have been proposed. See (\citeasnoun{ishizaka2011review}) for an account of the existing methods.  For example, the authors (\citeasnoun{alonso2006consistency}) have computed a regression of the random indices and proposed the following formula
\begin{equation}
\lambda_{\text{max}} < 1.17699 \, n - 0.43513 \,  ,
\end{equation}
where $n$ is the size of the pairwise comparison matrix, while (\citeasnoun{crawford1985note}) propose to use the \textit{Geometric Consistency Index} GCI
\begin{equation}\label{GCI}
\text{GCI} =  \frac{2}{(n-1)(n-2)}\sum_{i=1}^{n-1}\sum_{j=i+1}^{n}\left[\log\left(\frac{a_{ij}}{w_i/w_j} \right)\right]^2 \, .
\end{equation}
In the coming sections, we will make extensive use of the GCI for the computation of consistency of decision processes as we believe it is more apt to capture the propagation of inconsistencies.

\section{Methodology}\label{methodology}

In this section, we propose a methodology to evaluate R\&D performance. In particular, we discuss in detail how criteria and sub-criteria are to be chosen in our proposed method.  

\subsection{Criteria and sub-criteria in R\&D performance evaluation}\label{choosing criteria and subcriteria}

\subsubsection{Perspectives to measure R\&D performance}

Determining R\&D performances usually relies on the identification of indicators (or metrics) relative to some criteria (perspectives).  Giving the same importance to all indicators and/or criteria can lead to an oversimplification of the R\&D measuring process and this, in turn, may lead to misinterpretation to the actual performance of an R\&D project (\citeasnoun{salimi2018evaluating}).

Thus,  it is crucial to correctly identify criteria and sub-criteria and subsequently determine relative importance. The latter step can be carried out by asking a team of experts to make pairwise comparisons between alternatives for both perspectives (criteria) and indicators (sub-criteria).

Following the literature, for example (\citeasnoun{kaplan1996using}), (\citeasnoun{bremser2004utilizing}), (\citeasnoun{lazzarotti2011model}) (\citeasnoun{salimi2018evaluating}), we lay out the four perspectives which are relevant for measuring R\&D performance:
\begin{itemize}
	\item Internal Business perspective (IB)
	\item Innovation and Learning perspective (I\&L)
	\item Financial perspective (F)
	\item Network and Alliances perspective (N\&A)
\end{itemize}

Let us spell out what each perspective is about. The \emph{Internal Business perspective} refers to internal resources, such as technological capabilities or human resources, that influence directly the performance of a project.  The \emph{Innovation and Learning perspective} refers to the development of new skills as the result of project activities.  \emph{Financial perspective}, instead, aims at capturing financial aspects of a project, with a focus on financial sustainability of a project. Finally,  the \emph{Network and Alliances perspective} refers to the interaction with different partners, such as external companies involved in project activities and realization of the results.

The authors (\citeasnoun{salimi2018evaluating}) consider also the ``Customer perspective'', which refers to the extent that R\&D satisfies the needs of customers. In the following sections, we will be interested mainly in projects which do not involve customers.  Thus, we will stick with the four criteria identified above.

The four perspectives presented here will be the four criteria of our decision process.  Indicators, i.e. sub-criteria, will be associated with each of the criteria in a way that we now describe.

\subsubsection{Selection of Indicators}

Let us briefly outline the three steps we propose are to be taken in order to determine indicators for each criterion. These will be labeled Step 0, 1 and 2 and can be summarized as follows:
\begin{itemize}

\item Step 0: Selection of relevant raw data, i.e. the building blocks for the final indicators, through a questionnaire given to a team of experts. 
	
\item Step 1: Identification of the right indicators from data selected at Step 0 through a second questionnaire.
	
\item Step 2: Pairwise comparisons between perspectives (criteria) and indicators (sub-criteria) according to the AHP method described in the previous section with some modifications that we describe later.

\end{itemize}

More in detail, in Step 0 we prepare a list of \textit{parameters} (raw data) that will be used to identify the indicators for the decision process.  The list, an example of which is given in Section \ref{case study}, is submitted to a team of experts who are asked to identify the parameters that are usually available in the projects they are involved in. This step is necessary to understand which parameters, among the proposed ones, are more versed to capture a project performance. 

In Step 1,  we ask the same team of experts to build, out of the raw data selected at Step 0, the indicators for the different perspectives. In particular, each of the participants is asked to form a number of \textit{normalized} indicators for each perspective. For example,  jumping ahead to the example of R\&D performance evaluation given in Section \ref{case study}, if we think that the number of findings in a given project (each given in a publication or presented at a conference) in the shortest time is a relevant indicator for Innovation and Learning, then we might propose as indicator: \textit{\# of findings/total time of the project}.

If, for any reason, the experts think that some quantities do not need to be normalized and can stand on their own, they are allowed to choose no denominator. Finally, a set of indicators for each perspective is formed according to the consensus they received from the experts. 

In Step 2, the team of experts is eventually asked to form pairwise comparison matrices, both between all criteria and sub-criteria. 
Nevertheless, there is an important caveat. Differently from the original AHP method, we require no strict reciprocity: $a_{ij}$ 
should not be necessarily equal to $1/a_{ji}$, but small (and sporadic) deviations are allowed. The reason for introducing such an 
inconsistency is that we would like to develop a method capable to capture and bypass possible inconsistencies 
that often influence decision processes in R\&D performance evaluation.

\subsection{AHP for evaluating R\&D performance}\label{RD performance}

As it should be by now clear,  in our method, the criteria for R\&D performance evaluation are represented by the four perspectives mentioned in the last section, while indicators -- relative to each criterion -- are the sub-criteria. Different projects in an evaluation session make up the alternatives. In brief, the alternative which scores the biggest global priority will correspond to the most impactful -- as for the chosen criteria -- project for R\&D.

\subsubsection{Pairwise comparisons of perspectives and indicators}

Let us define the pairwise comparison matrix among criteria $C \in \mathbb{R}^{4 \times 4}$ in the following manner
\begin{equation}
	C = \begin{pmatrix}
	c_{11} &c_{12} &c_{13} &c_{14}  \\ 
	c_{21} &c_{22} &c_{23} &c_{24}  \\ 
	c_{31} &c_{32} &c_{33} &c_{34}  \\ 
	c_{41} &c_{42} &c_{43} &c_{44}  
	\end{pmatrix}\, .
\end{equation}
Of course, $c_{ii} = 1$ for $i=1, \dots, 4$. 
The priority vector $v^*$ of $C$ can be easily computed as the geometric mean over the columns of $C$, see eqn. \eqref{geometric mean}, 
\begin{equation} 
	v^*=\begin{pmatrix}
	\left(c_{11}\,c_{12}\,c_{13}\,c_{14}\right)^{\frac{1}{4}} \\ 
	\left(c_{21}\,c_{22}\,c_{23}\,c_{24}\right)^{\frac{1}{4}} \\ 
	\left(c_{31}\,c_{32}\,c_{33}\,c_{34}\right)^{\frac{1}{4}}\\ 
	\left(c_{41}\,c_{42}\,c_{43}\,c_{44}\right)^{\frac{1}{4}}\\ 
	\end{pmatrix}\, .
\end{equation}
It turns out to be useful to our purposes to normalize it in such a way the sum of its components is 1
\begin{equation} 
	v=\frac{v^*}{\sum_{i=1}^4 v_i^*}\, .
\end{equation}

In the same fashion, we can define the pairwise comparison matrix among sub-criteria $A^{(c)} \in \mathbb{R}^{m_c \times m_c}$,
\begin{equation}
	A^{(c)} = \left(a^{(c)}_{ij}\right) = \begin{pmatrix}
	a^{(c)}_{11} & \cdots  & a^{(c)}_{1m_c} \\ 
	\vdots & \ddots & \vdots \\
	a^{(c)}_{m_c1} & \cdots & a^{(c)}_{m_cm_c}\end{pmatrix} \, ,
\end{equation}
where $c$ is an index that labels the different criteria (in our case there is 4 of them). We can define, just as in the case of criteria, the priority vector $w^{(c)}$ for each $A^{(c)}$
\begin{equation} 
	w^{(c)*}=\begin{pmatrix}
	\left(a_{11}\,a_{12}\, \cdots \,a_{1m_c}\right)^{\frac{1}{m_c}} \\ 
	\left(a_{21}\,a_{22}\, \cdots \,a_{2m_c}\right)^{\frac{1}{m_c}}\\ 
	\vdots \\ 
	\left(a_{m_c 1}\,a_{m_c 2}\, \cdots \,a_{m_c m_c}\right)^{\frac{1}{m_c}}\\ 
	\end{pmatrix}\, ,
\end{equation}
and normalize it so that
\begin{equation} 
	w^{(c)}= \frac{w^{(c)*}}{\sum_{i=1}^{m_c} w_i^{(c)*}}\, .
\end{equation}

It turns out to be useful to repack the vectors $w^{(c)}$into a matrix $W \in \mathbb{R}^{4\times N_{\rm ind}}$, with  $N_{\rm ind} = \sum_c m_c$  the total number of indicators,  in the following fashion
\begin{equation} 
	W=\begin{pmatrix}
	w^{(1)\, T} & \rule[.5ex]{1em}{0.4pt} \, 0 \, \rule[.5ex]{1em}{0.4pt}  & \rule[.5ex]{1em}{0.4pt} \, 0 \, \rule[.5ex]{1em}{0.4pt} & \rule[.5ex]{1em}{0.4pt} \, 0 \, \rule[.5ex]{1em}{0.4pt} \\ 
	\rule[.5ex]{1em}{0.4pt} \, 0 \, \rule[.5ex]{1em}{0.4pt} & w^{(2)\, T} & \rule[.5ex]{1em}{0.4pt} \, 0 \, \rule[.5ex]{1em}{0.4pt} & \rule[.5ex]{1em}{0.4pt} \, 0 \, \rule[.5ex]{1em}{0.4pt} \\
	\rule[.5ex]{1em}{0.4pt} \, 0 \, \rule[.5ex]{1em}{0.4pt} & \rule[.5ex]{1em}{0.4pt} \, 0 \, \rule[.5ex]{1em}{0.4pt} & w^{(3)\, T} & \rule[.5ex]{1em}{0.4pt} \, 0 \, \rule[.5ex]{1em}{0.4pt} \\
	\rule[.5ex]{1em}{0.4pt} \, 0 \, \rule[.5ex]{1em}{0.4pt} & \rule[.5ex]{1em}{0.4pt} \, 0 \, \rule[.5ex]{1em}{0.4pt} & \rule[.5ex]{1em}{0.4pt} \, 0 \, \rule[.5ex]{1em}{0.4pt} & w^{(4)\, T} 
	\end{pmatrix}\, .
\end{equation}

We can now compute the global weight of the $i$-th indicator as
\begin{equation} 
	P_i = (v^T W)_i = \sum_{j=1}^4 v_{j} \, W_{ji}\, , \qquad i=1,\dots, N_{\rm ind}\, .
\end{equation}

Note that $\sum_{i = 1}^{N_{\rm ind}} P_i = 1$ in our normalization. When there is more than one expert the global weight vectors for each expert have to be combined
so to obtain a unique global weight $P^{\rm (group)}$. We will do this again by considering the geometric mean over the experts, i.e we employ the AIP (Aggregation of Individual Priorities) method rather than the AIJ (Aggregation of Individual Judgments), see (\citeasnoun{dong2010consensus}),
\begin{equation}\label{group priority}
	P^{\rm (group)}_i = \frac{\prod_{k=1}^{N_{\rm exp}} \left(P_i^{(k)}\right)^{\frac{1}{N_{\rm exp}}}}{\sum_{j=1}^{N_{\rm ind}}\prod_{k=1}^{N_{\rm exp}} \left(P_j^{(k)}\right)^{\frac{1}{N_{\rm exp}}}} \, ,
\end{equation}
where $k$ runs over the number of experts, $N_{\rm exp}$, and $P_i^{(k)}$ is the global weight vector of the $k$-th expert. 

\subsubsection{Evaluating R\&D performance}

Finally,  we need to find a way to determine the priority (or score) of each of the alternatives, i.e. different projects in our case. 

Each of the indicators in a given project can be measured, in general, by means of a standard scale.  For instance, ``time of a project'' (see Section \ref{case study}) can be easily extrapolated once we know the date of beginning and end of that given project.  So it seems natural, in order to compute the score of each project, to multiply the indicator-global-priorities by the corresponding R\&D measurement and, in fact, here lies the central point of our method.

Once we have determined the global weight of each indicator, we should multiply it by its ``performance'' parameter.  For instance, going back to the example of \textit{\# of findings/total time of the project} mentioned in the previous section, the higher this number is, in a given project, the better the project itself will perform in the final evaluation. This will ensure that the project, among those taken into considerations,  with the most performing indicators will be the most valuable for R\&D.

However, the alert reader has surely noticed that this can lead to a nonsense, as R\&D measurement are often dimensionful quantities and it makes no sense to sum them up. Thus, what we propose is to ``map'' each R\&D measurement to a dimensionless parameter lying in the range 0 to 1 using the empirical Cumulative Distribution Function (CDF).

We remind the reader that the CDF of a real-valued random variable $X$ is the function given by
\begin{equation}
F_X(x) = P(X \leq x) \, ,
\end{equation}
where $P(X \leq x)$ is the probability that the random variable $X$ takes on a value less than or equal to $x$.  Among its properties, we have that the CDF is a non decreasing function of its argument and right-continuous. In particular, if $X$ is a continuous random variable 
\begin{equation}
\lim_{x \rightarrow - \infty} F_X(x) = 0 \, , \qquad \lim_{x \rightarrow \infty} F_X(x) = 1 \, .
\end{equation}
In integral form the CDF can also be expressed as
\begin{equation}
F_X(x) = \int_{- \infty}^x f_{X}(t) \, \mathrm{d} t \, ,
\end{equation}
where $f_{X}(x)$ can be interpreted as a probability density function for the variable $X$.  
It is quite trivial to prove that for a continuous random variable $X$, 
the random variable $Y = F_X(X)$ has a standard uniform distribution.\footnote{If $X$ is a discrete 
	random variable, then its CDF is given by
	\begin{equation*}
		F_X(x) = \sum_{x_i \leq X} P(X = x_i) \, ,
	\end{equation*}
	where $P(X = x_i)$ is the probability for $X$ to attain the value $x_i$.
	Clearly, in this case the map $Y=F_X(X)$ does not yield a variable with standard uniform 
	distribution: The resulting variable is still discrete and $P(Y=y)=P(X=F_X^{-1}(y))$.
	However, if $X$ can take sufficiently many values, it can be approximately seen as 
	a continuous variable and also the aforementioned result approximately holds.   
}
Indeed, 
\begin{equation}
	\begin{split}
		F_Y(y) & = P(Y\le y) = P(F_X(X)\le y) \\
		 & =  P(X \le F_X^{-1}(y)) = F_X(F_X^{-1}(y)) = y\, .
	\end{split} 
\end{equation}

In practice, we map each R\&D measurement variable $X_i$ using the corresponding empirical CDF,
in place of the true unknown CDF,
so to obtain variables having an approximately uniform distribution in the range 0 to 1.


Thus, the final R\&D performance can be computed by means of the following formula\footnote{We have assumed throughout that the larger an indicator performance is the more it will contribute to R\&D performance, $S_{R\&D}$. It might very well be that exactly the opposite happens for a given indicator: the smaller an indicator is the better it is in terms of performance. In that case, it is enough to replace $F_{X_i}(x_i) $ by $1 - F_{X_i}(x_i) $.}
\begin{equation} 
	S_{\rm R\&D} =  \sum_{i=1}^{N_{\rm ind}} P^{(group)}_i F_{X_i}(x_i) 
\end{equation}

Note that $F_{X_i}(x_i) \leq 1$ for any $i = 1, \dots, N_{\text{ind}}$. Therefore, $S_{\rm R\&D} \leq \sum_{i=1}^{N_{\rm ind}} P^{(group)}_i = 1$. Thus we conclude that the R\&D performance for each project is always normalized to lie in the range 0 to 1:
\begin{equation}
0 \leq S_{\rm R\&D} \leq 1 \, .
\end{equation}

\subsection{Consistency of the method}\label{Consistency of the method}

In AHP we are asked to make comparisons between each pair among the alternatives.  Even though in ideal situations there would not be any inconsistencies,  in real situations our decisions are subject to judgmental errors and conflicting with each other to some extent. 

In the following we will stick with the assumption that error factors are log-normal distributed with 0 mean. Let us then proceed to estimate what the variance in a generic R\&D performance evaluation is going to be for us.

\subsubsection{Uncertainty in R\&D performance}

As just remarked, it is commonly accepted that inconsistencies are log-normal distributed. For example, (\citeasnoun{shrestha1991statistical}) found that for a pairwise comparison matrix of dimension $n$ the variance of the error $\sigma^2$ is well approximated by the formula \eqref{GCI}, that we report here for clarity,
\begin{equation} 
	{\sigma}^2 = \frac{2}{(n-1)(n-2)}\sum_{i=1}^{n-1}\sum_{j=i+1}^{n}\left[\log\left(\frac{a_{ij}}{w_i/w_j} \right)\right]^2 \, ,
\end{equation}
where $a_{ij}$ is the pairwise comparison matrix and $w_i$ the components of the corresponding priority vector.

In our case, at the level of the four criteria (the four perspectives mentioned in the previous section) we would find an error of the form
\begin{equation} 
	\sigma^{2} = \frac{1}{3}\sum_{i=1}^{3}\sum_{j=i+1}^{4}\left[\log\left(\frac{c_{ij}}{v_i/v_j}\right)\right]^2
\end{equation}
while for each of the sub-criteria we find 
\begin{equation} 
	{\sigma^{(c)}}^2 = \frac{2}{(m_c-1)(m_c-2)}\sum_{i=1}^{m_c-1}\sum_{j=i+1}^{m_c}\left[\log\left(\frac{a_{ij}^{(c)}}{w_i^{(c)}/w_j^{(c)}}\right)\right]^2
\end{equation}
where, again, $c$ is an index labeling each of the criteria and $m_c$ is the number of sub-criteria for the criterion $c$. In a similar fashion, in (\citeasnoun{eskandari2007handling}) it is argued that the variances associated with each local weight are given by
\begin{equation} 
	\sigma_{v_i}^{2} = \frac{15}{16}\left[\sum_{j=1}^{4} v_j^2 -v_i^2\right]\sigma^2 v_i^2
\end{equation}
for the case of the four criteria, while it is of the following form
\begin{equation} \label{uncertainties sub-criteria}
	\sigma_{w_i^{(c)}}^{2} = \frac{m_c^2-1}{m_c^2}\left[\sum_{j=1}^{m_c} {w_j^{(c)}}^2 -{w_i^{(c)}}^2\right]{\sigma^{(c)}}^2 w_i^2
\end{equation}
for the case of the sub-criteria. Note that we are assuming no correlation among different criteria or sub-criteria. In this way we can also repack the errors of eqn. \eqref{uncertainties sub-criteria} in the following $4 \times N_{ind}$ matrix
\begin{equation} 
	\sigma_W^2=\begin{pmatrix}
	\left(\sigma_{w^{(1)}}^2\right)^T & \rule[.5ex]{1em}{0.4pt} \, 0 \, \rule[.5ex]{1em}{0.4pt}  & \rule[.5ex]{1em}{0.4pt} \, 0 \, \rule[.5ex]{1em}{0.4pt} & \rule[.5ex]{1em}{0.4pt} \, 0 \, \rule[.5ex]{1em}{0.4pt} \\ 
	\rule[.5ex]{1em}{0.4pt} \, 0 \, \rule[.5ex]{1em}{0.4pt} & \left(\sigma_{w^{(2)}}^2\right)^T & \rule[.5ex]{1em}{0.4pt} \, 0 \, \rule[.5ex]{1em}{0.4pt} & \rule[.5ex]{1em}{0.4pt} \, 0 \, \rule[.5ex]{1em}{0.4pt} \\
	\rule[.5ex]{1em}{0.4pt} \, 0 \, \rule[.5ex]{1em}{0.4pt} & \rule[.5ex]{1em}{0.4pt} \, 0 \, \rule[.5ex]{1em}{0.4pt} & \left(\sigma_{w^{(3)}}^2\right)^T & \rule[.5ex]{1em}{0.4pt} \, 0 \, \rule[.5ex]{1em}{0.4pt} \\
	\rule[.5ex]{1em}{0.4pt} \, 0 \, \rule[.5ex]{1em}{0.4pt} & \rule[.5ex]{1em}{0.4pt} \, 0 \, \rule[.5ex]{1em}{0.4pt} & \rule[.5ex]{1em}{0.4pt} \, 0 \, \rule[.5ex]{1em}{0.4pt} & \left(\sigma_{w^{(4)}}^2\right)^T 
	\end{pmatrix}
\end{equation}

Given that we are interested in estimating the final error affecting $S_{\rm R\&D}$ for each of the projects, it is necessary to see how the uncertainties propagate. In particular,  the variance error for the global weight of an indicator (for each of the experts) is found to be
\begin{equation}\label{error on global indicators}
	\sigma_{P_i}^2 = \sum_{j = 1}^4 \left( \sigma_{v_j}^2 W_{ji}^2 + v_j^2 (\sigma_W)_{ji}^2\right) \, , \qquad i=1,\dots, N_{\rm ind}\, .
\end{equation}
Note that, in order to derive eqn.  \eqref{error on global indicators}, we assumed that the uncertainty affecting the criteria and sub-criteria are independent of each other. Finally, in order to estimate the error affecting the global weight of an indicator for the total group of experts we use the general formula (see for instance (\citeasnoun{bevington1993data}))
\begin{equation} 
	\sigma_{P_i^{(group)}}^2 = \sum_{l=1}^{N_{\text{exp}}} 
          \sum_{j = 1}^{N_{\text{ind}}} \left(\frac{\partial P^{(group)}_i}{\partial P^{(l)}_j}\right)^2\sigma_{P^{(l)}_j}^2 \, ,
\end{equation}
where the derivatives are easily computed from eqn. \eqref{group priority} to be
\begin{equation}
\frac{\partial P^{(group)}_i}{\partial P^{(l)}_j} =  \frac{ P^{(group)}_i \left( \delta_{ij} - P^{(group)}_j \right)}{N_{\rm exp} P_j^{(l)}} \, .
\end{equation}
Here $\delta_{ij}$ is the kronecker delta: $\delta_{ij} = 1$ if $i = j$ and 0 otherwise.  The uncertainty on the final outcome $S_{\rm R\&D}$ is easily evaluated to be (the $x$'s are assumed to have no associated statistical error)
\begin{equation} \label{consistency final}
	\sigma_{S_{\rm R\&D}}^2 =  \sum_{i=1}^{N_{\rm ind}} \sigma_{P^{(group)}_i}^2 F(x_i)^2  \, .
\end{equation}

	

\section{Application of the method and Results}\label{case study}

In this section, we apply our methodology to R\&D performance of 34 projects stored in the DPR\&DI (Digital Platform for R\&D and Innovation Projects).

The DPR\&DI is a PaaS (Platform as a Service) for the management of R\&D and industrial innovation projects. It allows to monitor in real time the progress of any project, the storage of information and sharing of data. It can also be used to create connections between the various parties involved in the innovation process, creating a shared space for collaboration that connects researchers, innovators, institutions and funding agencies\footnote{It has been developed by Idea-re S.r.l. under the grant delivered by the Umbria Region ``POR FESR 2014-2020. Asse I Azione 1.3.1.  Sostegno alla creazione e al consolidamento di start-up innovative ad alta intensità di applicazione di conoscenza e alle iniziative di spin-off della ricerca''}.

The data analytics algorithms used to extract information in terms of performance-monitoring indices and in relation to innovation-based development strategies are part of the REEDIA project\footnote{The REEDIA project has been developed by Idea-re S.r.l. under the grant delivered by the Umbria Region “POR FESR 2014 – 2020 - Asse I Azione 1.4.1 “Living Lab” nei capoluoghi di Provincia di Perugia e Terni “Sostegno all’individuazione di soluzioni innovative a specifici problemi di rilevanza sociale attraverso l’utilizzo di ambienti di innovazione aperta LivingLAB”.}.

We discuss in detail the various steps to find a project performance by applying the general procedure explained in the previous sections.

\subsection{Step 0}

First of all, we lay out the raw data (Step 0) that we reckoned were necessary to build meaningful indicators to evaluate the R\&D performance of the projects in the DPR\&DI.

Let us start off by giving all the quantities that we believe are relevant to characterize the magnitude of a project

%
%

\begin{center}
	\begin{tikzpicture}
		\node [mybox] (box){%
			\begin{minipage}{0.85\columnwidth}
				\begin{itemize}
					\renewcommand\labelitemi{--}
					\item Duration of the project
					\item Number of calls for tenders
					\item Number of partners involved in the project
					\item Number of project activities 
					\item Number of people involved in the project
					\item Number of people with an education appropriate for the given topic
					\item Time spent on the project
					\item Equipment usage time
				\end{itemize}	
			\end{minipage}
		};
		\node[fancytitle, right=20pt] at (box.north west) {\bfseries Project};
	\end{tikzpicture}%
\end{center}

Second, we believe the impact on R\&D is driven also by the amount of findings for a given project. Thus, we proposed to consider also the following quantities:
%
%
\begin{center}
	\begin{tikzpicture}
		\node [mybox] (box){%
			\begin{minipage}{0.85\columnwidth}
				\begin{itemize}
					\renewcommand\labelitemi{--}
					\item Number of findings (papers, books, conferences,  exhibitions, others)
					\item Number of papers for a given project
					\item Number of books for a given project
					\item Number of conferences attended to present a given result
					\item Number of exhibitions attended to present a given result
					\item Number of patents for a given project
				\end{itemize}	
			\end{minipage}
		};
		\node[fancytitle, right=20pt] at (box.north west) {\bfseries Findings};
	\end{tikzpicture}%
\end{center}

Moreover, it is crucial to have indicators measuring the total costs of a given project, especially in order to quantify the sustainability of the project itself. Thus, we introduce raw data also for detailing financial reporting: 
%
%
%
%
%
\begin{center}
	\begin{tikzpicture}
		\node [mybox] (box){%
			\begin{minipage}{0.85\columnwidth}
				\begin{itemize}
					\renewcommand\labelitemi{--}
					\item Total cost of the project
					\item Total cost of the project team 
					\item Total cost of equipment
					\item Total cost of external suppliers
					\item Total cost of consultants
				\end{itemize}	
			\end{minipage}
		};
		\node[fancytitle, right=20pt] at (box.north west) {\bfseries Financial Reporting};
	\end{tikzpicture}%
\end{center}
and financial support:
\begin{center}
	\begin{tikzpicture}
		\node [mybox] (box){%
			\begin{minipage}{0.85\columnwidth}
				\begin{itemize}
					\renewcommand\labelitemi{--}
					\item Grant eligible expenses
					\item Tax credit eligible expenses
				\end{itemize}	
			\end{minipage}
		};
		\node[fancytitle, right=20pt] at (box.north west) {\bfseries Financial Support};
	\end{tikzpicture}%
\end{center}

At this point,  a team of experts was asked to give a ranking of the raw data just given in order to form a coherent set of indicators.  In particular, this led us to Step 1, where raw data are combined to form the indicators, as explained in Section \ref{choosing criteria and subcriteria}.

%

\subsection{Step 1}

As already anticipated, a statistical analysis made over the experts' opinions has led to a set of indicators that can be used to evaluate R\&D performance. These are reported in Table \ref{tab indicators}.
\begin{table*}[t]
	\centering
	\caption{Indicators selected by the team experts consulted to evaluate R\&D performance.}
	\label{tab indicators}
	\begin{tabular}[t]{ll}
		\toprule
		\textbf{Perspective} & \textbf{Indicators}\\
		\midrule
		\multirow{5}{*}{\textbf{Internal Business Perspective}}  &
		Number of findings / Cost of the project    \\ 
		& Number of people in the project / Project duration  \\ 
		& Grant eligible expenses  \\ 
		& Time spent on the project / Number of people involved \\ 
		& Time spent on the project / Number of activities  \\ 
		\midrule
		\multirow{6}{*}{\textbf{Innovation and Learning perspective}} &
		Number of papers / Number of people in the project \\
		& Number of books / Number of people in the project \\
		& Number of patents / Total cost of the project\\
		& Number of findings / Duration of the Project \\
		& Number of papers / Total cost of the project \\
		& Number of findings /  Time spent on the project		\\
		\midrule
		\multirow{6}{*}{\textbf{Financial perspective}} &
		Total cost of the team / Total cost of the project\\
		& Total cost of the suppliers / Total cost of the project \\
		& Total cost of equipment / Total cost of the project \\
		& Grant eligible expenses / Total cost of the project \\
		& Number of patents / Total cost of the project \\
		& Time spent on the project / Total cost of the project \\
		\midrule
		\multirow{4}{*}{\textbf{Alliances and Networks perspective} } &
		Number of partners \\
		& Number of partners  / Time spent on the project \\
		& Number of project activities  / Total cost of suppliers  \\
		& Number of patents / Number of suppliers  \\
		\bottomrule
	\end{tabular}
\end{table*}%
As we can see, there are 5 indicators for the Internal and Business perspective, 6 for Innovation and Learning, 5 for the Financial perspective and 4 for Alliances and Network perspective. We have thus created a layer of 20 indicators (sub-criteria), each associated with a given perspective (criterion). This, along with the 34 project considered in this study, makes up the basic AHP structure in the R\&D performance evaluation.

\subsection{Step 2}

We are now ready, as for Step 2, to compute the R\&D performance for the 34 selected projects using formulas spelled out in Section \ref{RD performance}.

\begin{figure*}[!h]
	\centering
	\begin{subfigure}[t]{0.49\textwidth}
		\centering
		\includegraphics[width=\textwidth]{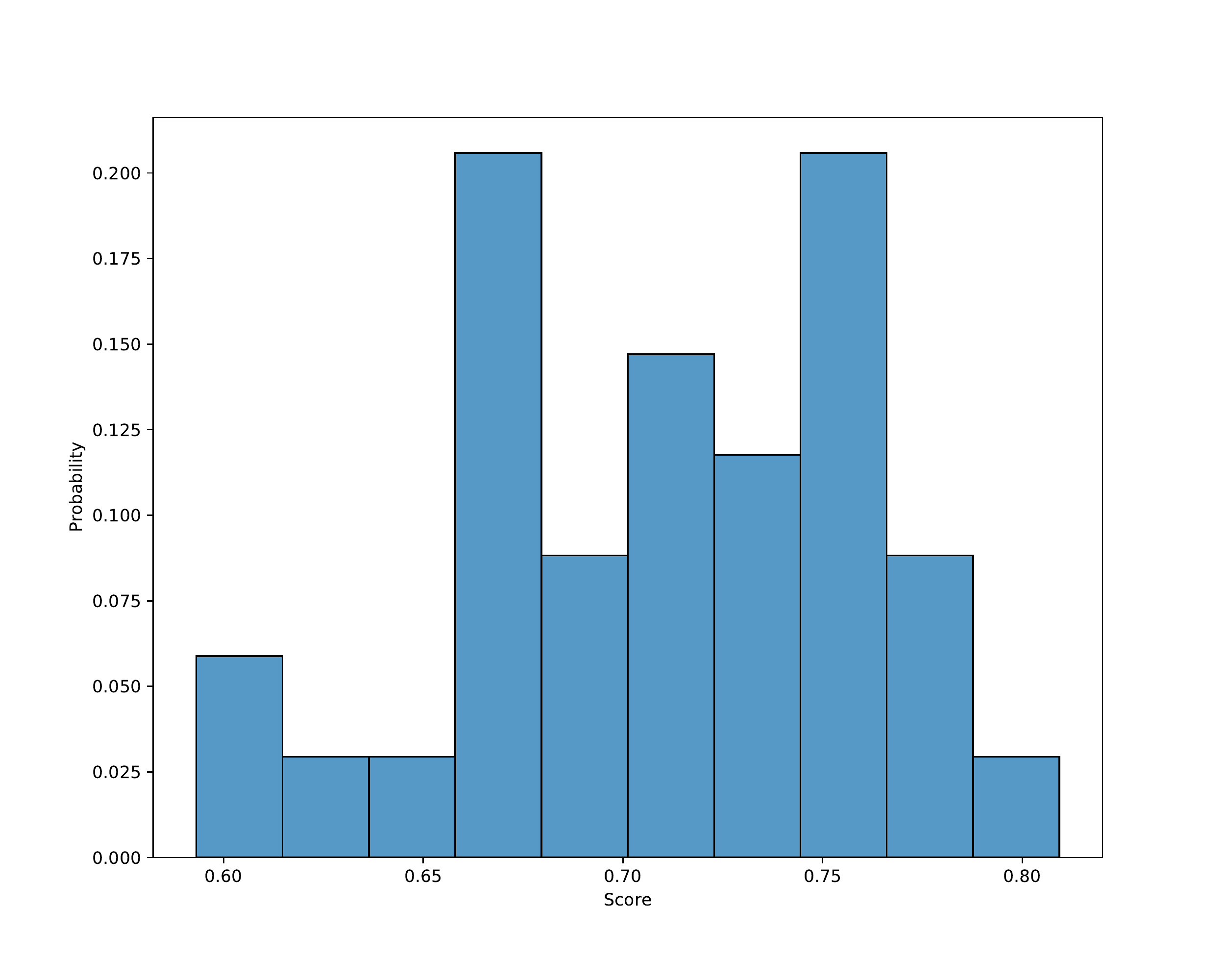}
		\caption{\footnotesize{Histogram of the score distribution. On the $y$-axis we have the relative probability of finding a given score ($x$-axis)}.}
		\label{fig:score_distribution}
	\end{subfigure}
	\hfill
	\begin{subfigure}[t]{0.49\textwidth}
		\centering
		\includegraphics[width=\textwidth]{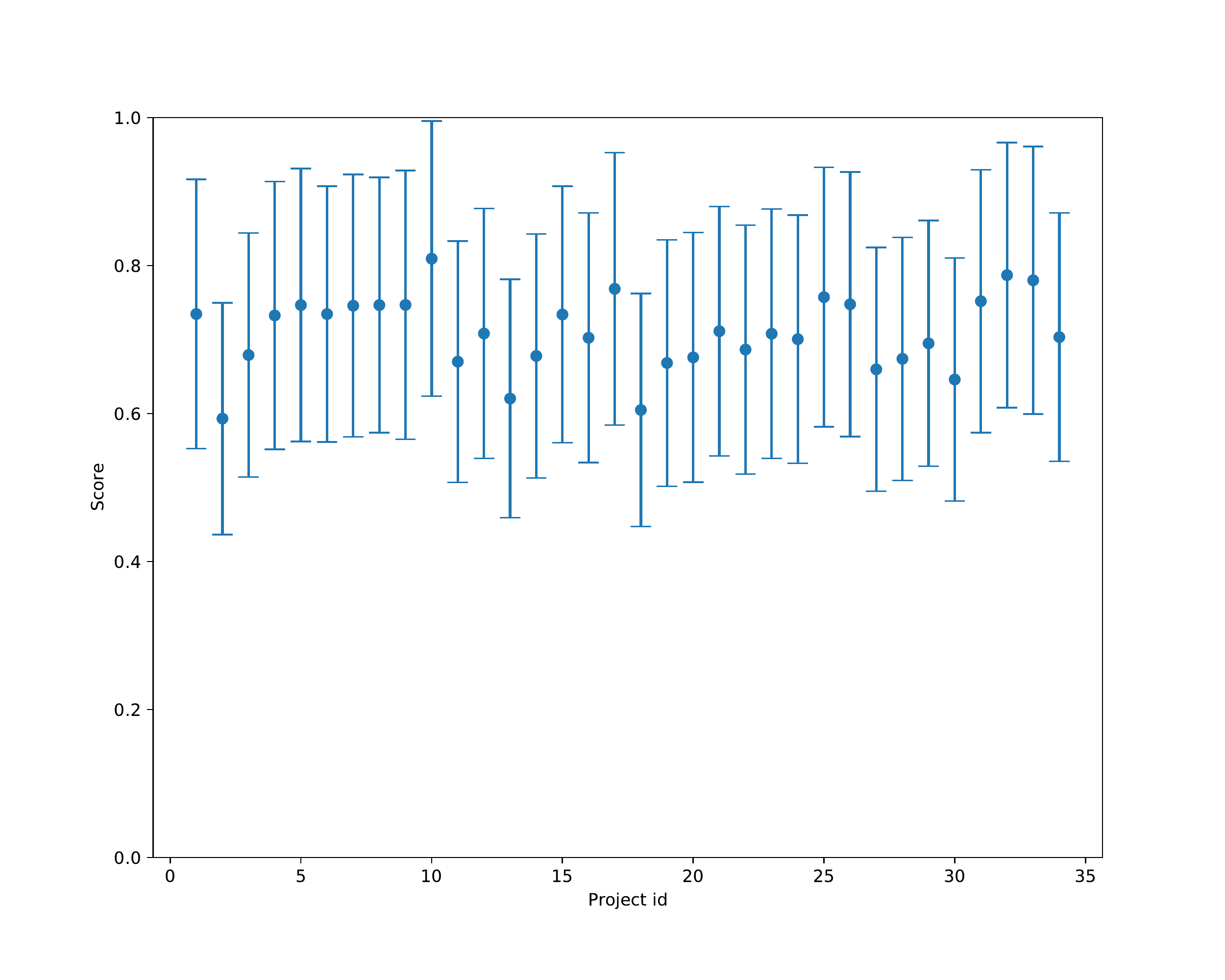}
		\caption{\footnotesize{Scores with error bars. }}
		\label{fig:scores_errors}
	\end{subfigure}
	\caption{Score distribution for 34 projects stored in the DPR\&DI.}
	\label{fig:scores}
\end{figure*}

In particular, the distribution for the R\&D performance scores is depicted in Fig.~\ref{fig:score_distribution}. We can see that the distribution is quite uniform, and all scores lie (approximately) in the range 0.6 to 0.8 (remember that the $S_{R\&D}$ is normalized to be in the range 0 to 1).

As for the consistency of our results we can employ the formula \eqref{consistency final}. Scores with errors are shown in Fig.~\ref{fig:scores_errors}. 

As we can see, the $\sigma^2$ on any given project is quite significant, making it hard to identify precisely which project performs best in this particular analysis. This is essentially due, as we would expect,  to the degree of inconsistency allowed when forming the pairwise comparisons. What could be nice to do is to compute the probability of inversion of two given projects in the final ranking. We leave issues like this for future studies.

\section{Discussion and Conclusions}\label{conclusions}

In this paper we considered a new approach to determine R\&D performance based on the group-AHP method.  As explained thoroughly in the main text, the AHP method is a powerful method that allows to quantify relative weights of criteria in a decision problem. In particular, any decision process is suitably decomposed into a hierarchy of sub-problems that are usually rather easy to deal with.

In this paper the decision process of the AHP method corresponds, roughly speaking, to determining which among a list of R\&D projects has the best performance according to a number of criteria (perspective) and sub-criteria (indicators) selected by a team of experts.

The need for a systematic and quantitative analysis of the performance of R\&D projects relies on the fact that, nowadays,  R\&D is one of the most significant determinants of the productivity and growth of companies, organizations, governments etc. Thus, it has become somewhat crucial to have at our disposal an intuitive, easy, efficient yet systematic and analytical method to quantify R\&D performance.

More in detail, we started off in Section~\ref{Basics of the AHP method} by describing the basics of AHP method as originally developed by Saaty, outlining all the important steps to follow in a decision process in order to determine the best among a set of alternatives. 

In Section \ref{methodology} we laid out the general procedure of our proposed method in order to define the basic AHP structure for R\&D performance evaluation.  As we have seen in the main text, this is essentially based on a set of questionnaires handed to a team of experts who are asked, through a number of steps, to define a consistent hierarchical structure of the AHP-based method. Then we gave more mathematical details on how a quantitative evaluation of R\&D performances and relative inconsistencies can be carried out. Finally, in Section \ref{case study} we presented an example of our method for the case of a number of projects stored in the DPR\&DI platform. 

We believe that our results might have important implications for those companies, organizations and public administrations interested in determining R\&D performance. First of all,  we provided a method for a firm to make comparisons between its R\&D projects. In this way managers are facilitated in understanding which project is more deficient and in which area (perspective) or even in formulating more effective strategies to improve the R\&D performance of low-scoring projects according to their own objectives.  Second,  our method offers a way of comparing a company's R\&D global performance to the performance of other firms. 

To sharpen our work further, it could be interesting to study and quantify the compatibility of the different experts (i.e.  how far off they are with respect to one another) involved in the decision process. 
See for instance (\citeasnoun{aguaron2019ahp}).  Another interesting direction might be that of gathering data from different experts (eqn. \eqref{group priority}) 
using a weighted geometric mean. For example, we could set up a computation where the more consistent an expert has been in writing down pairwise comparison matrices, 
the more weight she/he will have in the computation of priorities.  Moreover, it would be interesting to find a way of discussing more perspectives 
than those considered in this paper (Internal business, Innovation and Learning, Financial and Network and Alliances perspectives). 
In this way, we may hope to build a more general method suitable to many more organizations. We hope to tackle all these problems in the near future.

\section*{Acknowledgments}

It is a great pleasure to thank the Italian Association of University Professors of Mechanical and Thermal Measurement for its support during the realization of the present paper.



\balance
\bibliography{harvard}
\end{document}